\documentclass[a4paper,12pt]{article}
\usepackage{epsfig,float,wrapfig,amsmath,graphicx}
\begin{document}

{\Large
\begin{center}
{\bf Calculations of $K^\pm$ and  $\phi$ Production in Near-Threshold 
Proton-Nucleus Collisions} 
\end{center}
}

\begin{center}
H.W. Barz$^1$, B. K\"ampfer$^1$, L. Naumann$^1$, Gy. Wolf$^2$ 
and M. Z\'et\'enyi$^2$\\[10mm] 
\center \begin{minipage}[b]{130mm}
$^1$ Forschungszentrum Rossendorf, Pf 510119, 01314 Dresden, Germany\\ 
$^2$ KFKI Budapest,  POB 49, H-1525 Budapest, Hungary 
\end{minipage}
\end{center}
\vspace*{10mm}

{ $K^\pm$ and $\phi$ meson production 
in proton-nucleus (pA) collisions has been 
calculated within a BUU transport model. It is shown 
that the nucleon-hyperon strangeness transfer channel is essential. The role of 
three-body reactions has been investigated within the medium.
The target 
mass dependence of $\phi$ production is predicted to give  important
information on the in-medium properties of all three mesons.
}
\vspace*{8mm}


\section{Introduction}\label{intro}

Heavy-ion reactions and hadron-nucleus reactions are an
important tool to investigate in-medium properties of produced 
particles. Heavy-ion collisions probe the properties at
high densities while hadron beams interact with matter of normal nuclear
density $n_0$. At threshold the production rates 
depend sensitively on the potentials which reflect the interaction with
the ambient medium. If the reaction mechanism would be sufficiently known
the potential can be fixed by comparison with data. 
Therefore, we will concentrate in this contribution on the different elementary
reactions that lead to particle production near the respective thresholds
in free nucleon-nucleon collisions.

An increasing number of sophisticated studies of  $K^+$  production has lead to 
sufficient experience for their theoretical description. 
Cross sections
as estimated in refs. \cite{tsushima1,tsushima2} have 
successfully been applied and 
a weak repulsive potential being roughly  proportional to the 
nuclear density $n$ of about 25 $n/n_0$ MeV has been extracted
\cite{cassing0}.
This 
knowledge is very important for  understanding the production of $K^-$
which in threshold region are mainly produced in secondary collisions 
of nucleons and pions with hyperons ($Y$) that has been created together with 
$K^+$ mesons.
Nevertheless, the mechanism for production of
the rarer mesons $K^-$ and $\phi$ is not well understood.

Our contribution is organized as follows. 
In section \ref{NY} we investigate the role of the $NY\to NNK^-$
channel the contribution of which has been largely underestimated in the
past. The role of three-body collisions we discuss in subsection 
\ref{threebody}
in connection with $\phi$ production. In subsection \ref{Amass} 
we investigate how the in-medium
properties of the $\phi$ meson influence the production rates as a
function of the target mass.

\begin{figure}[htb]
\begin{center}
\includegraphics[width=100mm]{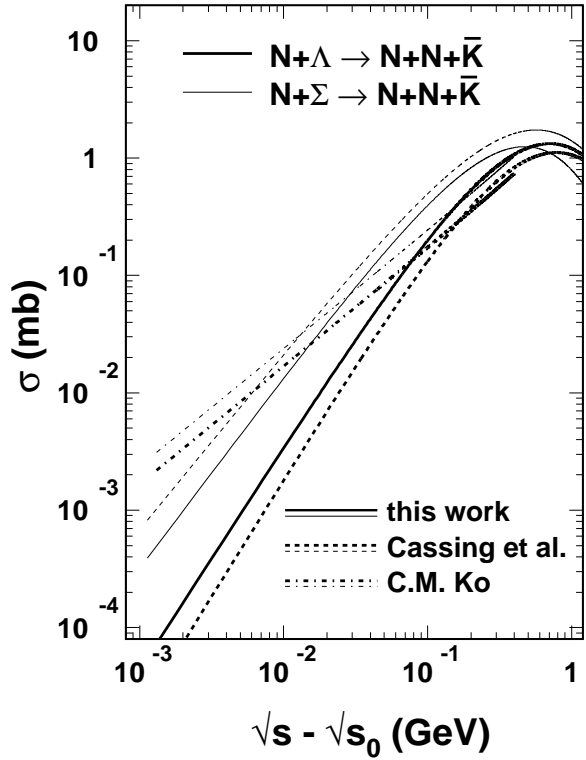}
\end{center}
\vspace*{-1.7cm}
\caption{\small Comparison of isospin averaged production cross sections 
for antikaon production as a function of excess energy obtained 
by various models. 
The thick (thin) lines refer to $N\Lambda$($N\Sigma$) collisions, 
whereas the line pattern refer to different authors
\cite{ko1,cassing1}.
  }
\label{fig1}
\end{figure}

\section{The Role of the NY Channel in pA Collisions}\label{NY}  
 
Strangeness transfer reactions have a comparably large cross section
for $K^-$ production 
since no strange quark pairs have to be created. 
In heavy-ion collisions (AA) the reactions 
$\pi Y$ and $NY$ are relevant. In  the former reaction  two
secondary particles have to be created as initial state.
This can easily be achieved in 
AA collsions. However, 
in pA collisions the probability that the incoming proton alone can create
both particles is small and the latter reaction should become more important 
despite its higher threshold. In ref.~\cite{barz1} we have investigated its
contribution in pA collisions. First estimates of the experimentally unknown
cross section were done in ref.~\cite{ko1} and later in ref.~\cite{cassing1}.
We have extended these calculations using the experimentally known $K^-N$
cross sections. Our results fairly agree with the foregoing estimates
as can be seen in Fig.~\ref{fig1}.

\begin{figure}[htb]
\begin{center}
\includegraphics[width=100mm]{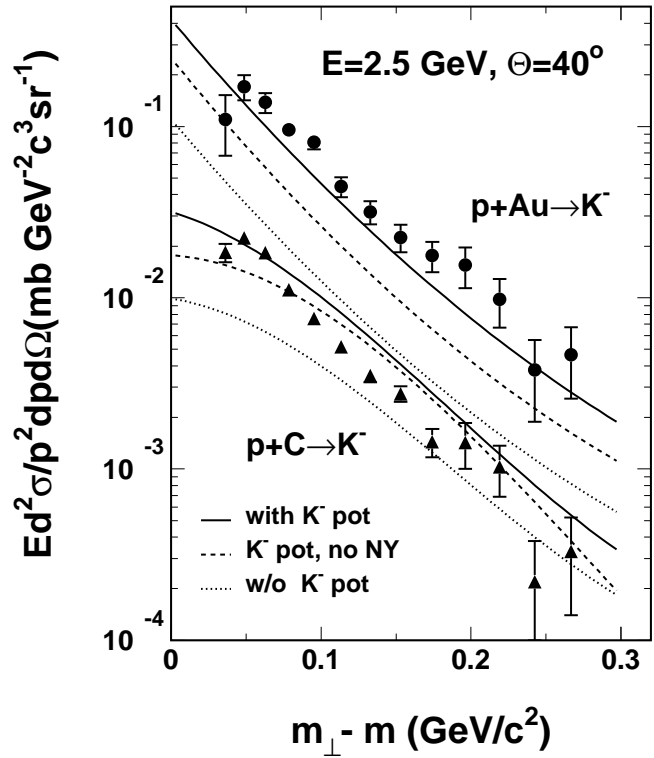}
\end{center}
\vspace*{-1.7cm}
\caption{\small Comparison of measured $K^-$ production cross sections to 
different calculations within the BUU model. The full (dotted) lines show
calculations with (without) using an antikaon potential of -120 $n/n_0$ MeV
while the dashed curves indicate calculations without the contribution
of the nucleon-hyperon channel.
  }
\label{fig2}
\end{figure}

Now we study the role of the NY $\to$ NNK$^-$ reaction
in pA collisions. 
 The \mbox{NY channels} are incorporated 
 into a transport model calculation which is based
 on the Boltzmann-\"Uhling-Uhlenbeck (BUU) equation \cite{wolf1}.
 In Fig. \ref{fig2} we present the differential cross sections
 at a laboratory angle of 40$^\circ$ for collisions of protons 
  at 2.5 GeV beam energy with
 $^{12}$C and $^{197}$Au.
 The dotted lines are calculated without using potentials for the
 kaons and antikaons. These calculations
 underestimate clearly the measured data \cite{KAOS6}.
 The attractive antikaon potential of -120$n/n_0$ MeV 
 in addition with the \mbox{NY channels} leads to an increase of the
 cross section as shown by the full lines. 
 The dashed curves depict results where the $NY\to K^-$ channels
 has been excluded.
 Disregarding these channels
 the cross section diminishes by about 40\% in pAu collisions.
 This shows the importance of the $NY\to K^-$ channels when
 one intends to determine the K$^-$ potential.
 For the
 light C target
 the influence is much smaller as the hyperons have a smaller
 chance to collide with further nucleons before leaving the reaction zone.

\begin{figure}[htb]
\vspace*{-0.1cm}
\begin{center}
\includegraphics[width=120mm]{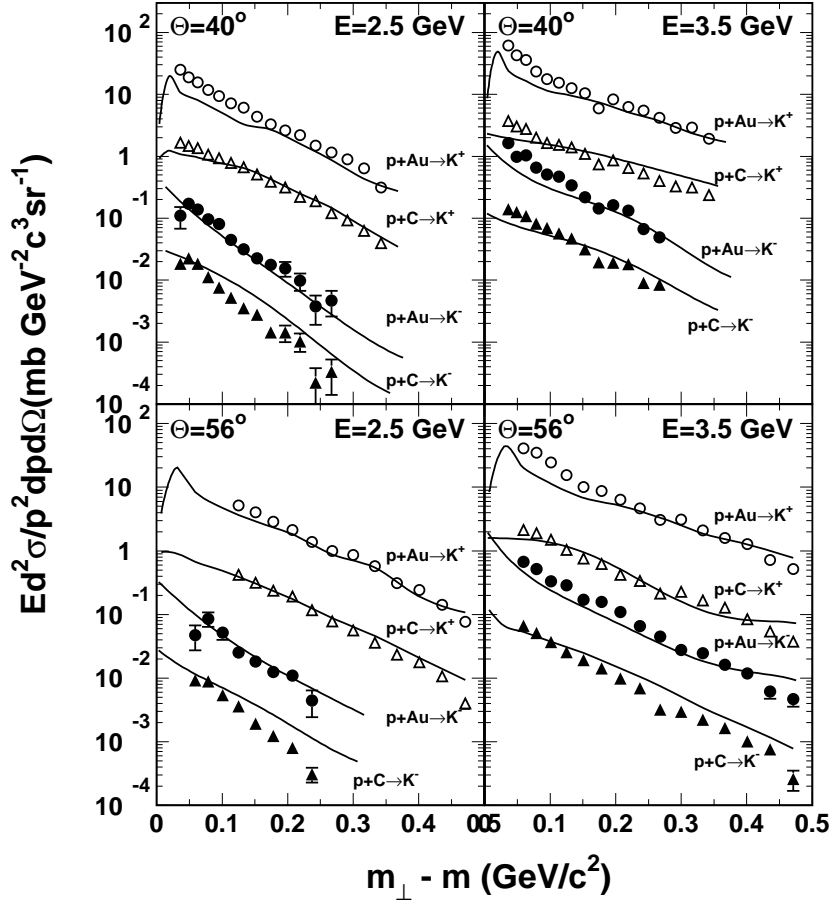}
\end{center}
\vspace*{-1.2cm}
\caption{\small
Invariant differential K$^\pm$ cross section as a function of
the 
transverse kaon mass at 2.5 and 3.5 GeV proton beam energy.
The solid lines refer to our
calculations including the NY channels for the antikaon production.
Data are taken from ref.~\cite{KAOS6}.
}
\label{fig3}
\end{figure}

Finally, we compare in Fig.~\ref{fig3} our calculations
for $K^+$ and  $K^-$ production
with data obtained by the KaoS collaboration \cite{KAOS6}
for proton-nucleus collisions at bombarding energies of 2.5 GeV and
3.5 GeV on C and Au targets and $K^\pm$ meson emission angles of
$40^{\circ}$ and $56^{\circ}$.
The kinetic beam energy ${T_\mathrm{kin}=2.5}$~GeV is close to the
production threshold in nucleon-nucleon
collisions.
In the calculations we  use the parameters of
ref.~\cite{tsushima1,tsushima2} and
obtain $K^+$ cross sections which fairly well agree with 
the data for for both targets and angles.

\section{$\phi$ Production}\label{Phi}  

\subsection{The Role of Three-Body Collisions}\label{threebody}

Now we study whether three-body
collisions could remarkably contribute to
the production of $\phi$ mesons.
At threshold such a mechanism could be dominant
because less bombarding energy is required if the
projectile nucleon interacts with two target nucleons instead with
a single one.  For two nucleons the threshold reduces from
2.6 GeV to 1.8 GeV.

We describe three-body processes by Feynman diagrams with propagators
representing the propagation of intermediate particles \cite{barz2}. 
There are subspaces in the phase space,
where the intermediate particles move on shell.
Such processes are described within the current transport models
by sequential two-step processes which also allow to accumulate
the energy of several nucleons. It is our aim to separate the 
on-shell and off-shell propagation and 
relate  them to the standard treatment as sequential
two-step processes.

To determine the on-shell contribution we define the genuine three-body
cross section as the difference of the total cross section
and all possible two-step cross sections. In ref.~\cite{barz2}
it has been shown that the difference always has a finite value
even when using  the undressed  propagators. 
In a heavy-ion collisions the intermediate particles interact with 
the medium which can be approximately described by adding the
collisional width $\Gamma$ to the mass 
$m_c$ of the intermediate particle. 
In this case the three body term is 
calculated by the difference
\begin{equation}
\sigma_{\rm three} = C\int d{\bf L}_{\rm inv} \bigg( |{\cal{T}}|^2
     \;-\; \sum_{c} |\hat{T}_c|^2 \delta(p_c^2-m_c^2) \,
                  \frac{\pi}{\Gamma m_c}\bigg)\,.    \label{threemore}
\end{equation}
Under the phase-space integral the sum runs over all open channels $c$ for
 two-step collisions, where resonances
 are  included as intermediate particles as long as their decay channels are open. 
The full T matrix of the process is denoted by ${\cal{T}}$ 
while $T_c$ is the product
of the two T matrices describing the subprocesses of the two steps of the 
production \cite{barz2}.

\begin{figure}[htb]
\vspace*{-1.1cm}
\begin{center}
\includegraphics[width=100mm]{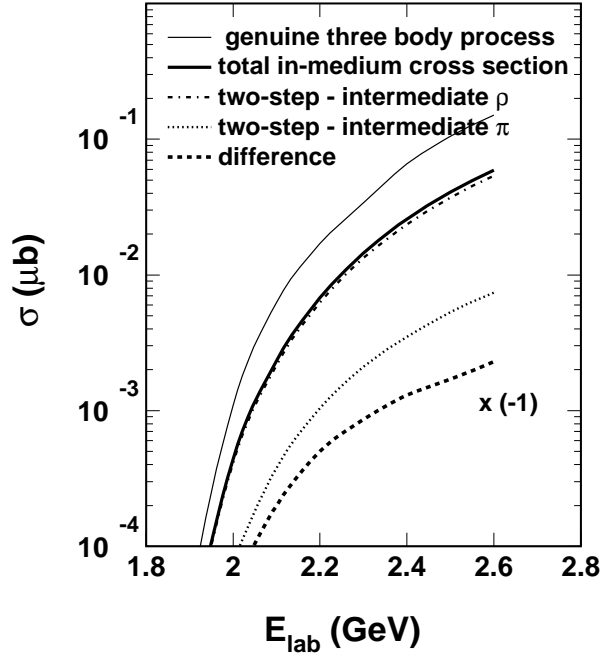}
\end{center}
\vspace*{-1.7cm}
\caption{\small Cross sections for $\phi$ production by a proton colliding
with two protons at rest. For details see text.
  }
\label{fig4}
\end{figure}

In Fig.~4 we show the result for the $\phi$ production for 
a proton colliding with a two protons at density $n_0$.
The three-body
cross section is calculated without in-medium effects and 
is presented in Fig.~\ref{fig4} by the thin upper line.
However the total cross section
(thick full line) is much smaller 
if the rescattering
effects of the intermediate particles are taken into account.
(We have used an in-medium cross section of 25 mb 
for both pions and $\rho$ mesons.)
The sum of the two corresponding two-step processes (dashed and dot-dashed lines)
with an intermediate $\rho$ meson and pion, respectively, 
has nearly the same value as the total cross section.
In the case considered here the two-step processes
slightly overestimate the cross section (see thick dashed line).
Thus it turns out that genuine three-body processes are not important
as most of the reaction rate can be evaluated via consecutive two-body
reactions. 
However, it is necessary that all intermediate particles
which can be transferred to the mass shell are treated properly
within the transport code.

\subsection{Target Mass Dependence of $\phi$ Production }\label{Amass}

The study of $\phi$ meson production in pA collisions provides an
independent test of the in-medium kaon and $\phi$ potentials. In case
of a strong attractive $K^-$ potential and moderate $K^+$ and $\phi$
potentials the mean proper lifetime of $\phi$ mesons decreases
from the vacuum value of about 50 fm/$c$ to an order of
magnitude smaller value in normal nuclear matter. This effect is
mainly caused by the attractive $K^-$ potential. Therefore, $\phi$
mesons created in a pA collision have a large probability to decay
into kaon pairs inside the nucleus. These kaons may rescatter
before they leave the nucleus, therefore the kinematic information
needed to reconstruct the $\phi$ meson might be lost. This effect is
expected
to be bigger for a larger nucleus. Therefore, studying the mass dependence
of the number of $\phi$ mesons reconstructed from the K$^+$K$^-$
channel is a suitable probe for studying the in-medium broadening
of the $\phi$ meson.

In the following we present results \cite{barz3} obtained for different 
$K^-$ potentials of the form $V_{K^-} = -(v+w*exp(-sp_K/{MeV}))$ MeV
with $p_K$ being the kaon momentum. We use 
four different parameter sets $(v,w,s)$ of increasing strength: 
(i) (0,0,0), (ii) (70,0,0), (iii) (55, 130, 0.0025) and (iv) (150,0,0).

\begin{figure}[htb]
\vspace*{-1.1cm}
\begin{center}
\includegraphics[width=100mm]{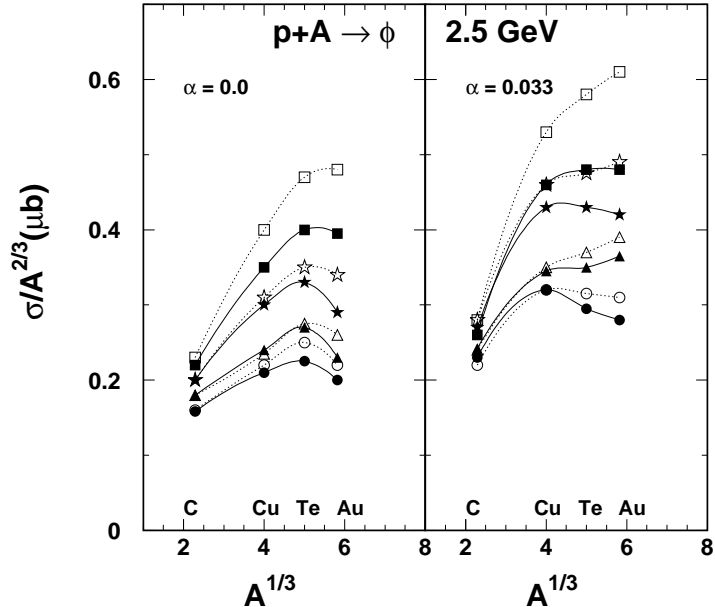}
\end{center}
\vspace*{-1.0cm}
\caption{\small Rescaled cross sections for $\phi$ meson production as
a function of the target mass for different $K^-$ potentials.
The left panel shows
the result using  the vacuum $\phi$ mass while on the right hand
side  the mass is diminished using 
$\alpha=0.033$. The cross sections are
 reconstructed from electron pairs (open symbols) and
 from kaon pairs (full symbols), respectively.
 Symbols denote $K^-$ potentials i(see text) used: squares (i), 
 stars (ii), triangles (iii) and circles (iv).
  }
\label{fig5}
\end{figure}

The comparison of the left and the right hand part of Fig.~\ref{fig5}
shows the effect of the change of the  mass of the $\phi$ meson.
We have assumed that the relative mass change is proportional to $\alpha n/n_0$.
Without a K$^-$ potential the $\phi$ production cross
section increases faster than the geometrical cross
section as a function of the target mass number $A$. 
For masses below the copper mass the cross
section is roughly proportional to the mass number.
This is because the $\phi$ mesons are predominantly created
in two-step processes by secondary $\pi$ and $\rho$ mesons.
For larger nuclei the increase is moderate because $\phi$ mesons
get absorbed due to $\phi N\to\pi N$ reactions. This effect is especially 
seen on the  right hand 
side of  Fig.~\ref{fig5} where the $\phi$ mass is reduced 
and the decay inside the nucleus is hindered.
$\phi$ meson production in pA collisions can also be studied via the
dilepton-decay channel $\phi\to e^+e^-$. Since electrons are marginally
influenced by the nucleus 
the $\phi$ cross sections  extracted from their invariant mass 
are somewhat larger.

\section{Conclusions}\label{concl}
In summary we have reviewed our transport model studies of $K^\pm$ and
$\phi$ meson production in proton-nucleus collisions near threshold.
Inclusive $K^\pm$ production requires strong strangeness transfer 
channels $NY\to NNK^-$ 
and a mildly repulsive $K^+$  and a strongly attractive 
$K^-$ potential. The correlated $K^+K^-$ yields are sensitive to in-medium effects
of all three mesons which may be accessible in a precision measurement
of the $K^+ K^-$ pairs with a invariant mass 
corresponding to the $\phi$ meson. The tight coupling
of $K^-$ and $\phi$ production rates has been emphasized in ref.~\cite{kotte}
for heavy-ion collisions. The systematic study of meson production
of open and hidden strangeness in proton-nucleus and  
heavy-ion collisions yields complementary information on the importance
of the various elementary reaction channels.
 
\section*{Acknowledgments}
This work is supported  by the German ministry BMBF, the GSI-FE fund,
the German DAAD scientific exchange program, the National Fund for Scientific
Research of Hungary T047347 , the
S\"achsische Staatsministerium f\"ur Wissenschaft und Kunst,
and the Bergen Computational Physics Laboratory.

\vfill\eject
\end{document}